\documentclass[fleqn,12pt,twoside]{article}
\usepackage[headings]{espcrc1}

\readRCS
$Id: espcrc1.tex,v 1.2 2004/02/24 11:22:11 spepping Exp $
\usepackage{graphicx}
\usepackage[figuresright]{rotating}

\newcommand{\beq}{\begin{equation}}
\newcommand{\eeq}{\end{equation}}
\newcommand{\bqa}{\begin{eqnarray}}
\newcommand{\eqa}{\end{eqnarray}}

\newcommand{\AmS}{{\protect\the\textfont2
  A\kern-.1667em\lower.5ex\hbox{M}\kern-.125emS}}

\title{The role of quark mass in cold and dense pQCD and quark stars}

\author{Eduardo S. Fraga\address{Instituto de F\'\i sica, 
Universidade Federal do Rio de Janeiro\\ 
C.P. 68528, Rio de Janeiro, RJ 21941-972, Brazil}}
       
\begin{document}

\maketitle

\begin{abstract}
For almost twenty years the effects of a nonzero strange 
quark mass on the equation of state of cold and dense QCD 
were considered to be negligible, thereby yielding only minor 
corrections to the mass-radius diagram of compact stars. 
By computing the thermodynamic potential to first order 
in $\alpha_s$, and including the effects of the renormalization 
group running of the coupling and strange quark mass, we show 
that corrections can be of the order of 25$\%$, 
and dramatically affect the structure of compact stars. 
\end{abstract}

\vspace{0.75cm}

Compact stars provide the most promising ``laboratory'' to 
constrain the equation of state for strong interactions.
In this work we compute the thermodynamic potential for 
cold quark matter with two light (massless) flavors, corresponding to the 
up and down quarks, and one massive flavor, corresponding to the 
strange quark, in perturbation theory to first order in $\alpha_s$ in the 
$\overline{\hbox{\scriptsize MS}}$ scheme \cite{Fraga:2004gz}. 
In this fashion, we can easily 
include modern renormalization group running effects 
for both the coupling and the strange quark mass. We find that 
the corrections due to the running nonzero mass are sizable, and should 
not be neglected in the evaluation of thermodynamic quantities. Solving the 
Tolman-Oppenheimer-Volkov (TOV) equations using our equation of state 
in the presence of electrons, we show that a running strange quark mass 
dramatically modifies the mass-radius diagram for quark stars, even at 
first order in $\alpha_s$.

The thermodynamic potential for cold QCD in perturbation theory to 
$\sim\alpha_s^2$ was first computed a long time ago 
\cite{Free,Baluni,Toimela}. Nevertheless, 
the original approach to quark stars \cite{wit,bag} made use of 
the bag model with corrections $\sim\alpha_s$ from perturbative QCD to 
compute the thermodynamic potential. 
In the massless case, first-order corrections cancel out in the 
equation of state, 
so that one ends up with a free gas of quarks modified only 
by a bag constant.  
Finite quark mass effects were then estimated to modify the equation of 
state by less than $5\%$ and were essentially ignored. 

A few years ago, corrections $\sim\alpha_s^2$ with a modern definition 
of the running coupling constant were used to model the non-ideality 
in the equation of state for cold, dense QCD with massless quarks 
\cite{Fraga:2001id,Fraga:2001xc}. 
This approach can be compared to treatments that resort to 
resummation methods and quasiparticle model descriptions 
\cite{andre,tony,andersen,paul}. Remarkably, these different frameworks seem 
to agree reasonably well for $\mu >> 1~$GeV, and point in 
the same direction even for $\mu \sim 1~$GeV and smaller, 
pushing perturbative QCD towards its border of applicability. 

Even the most recent QCD approaches mentioned above generally neglected 
quark masses and the presence of a color superconducting gap as compared 
to the typical scale for the chemical potential in the 
interior of compact stars, 
$\sim 400~$MeV and higher. However, it was recently argued that both effects 
should matter in the lower-density sector of the equation of 
state \cite{alford}. 
In fact, although quarks are essentially massless in the core of quark stars, 
the mass of the strange quark runs up, and becomes comparable to the typical 
scale for the chemical potential, as one approaches the surface of the star. 

In what follows, we present an exploratory analysis of the effects 
of a finite mass 
for the strange quark on the equation of state for perturbative QCD 
at high density, 
leaving the inclusion of a color superconducting gap in this framework 
for future investigations. To illustrate the effects and study the 
modifications in the 
structure of quark stars, we focus on the simpler case of first-order 
corrections. 
Results including full corrections $\sim\alpha_s^2$, as well as 
technical details 
of the calculation at each order and renormalization, will be 
presented in a longer 
publication \cite{next}.

The leading-order piece of the thermodynamic potential of QCD for one massive 
flavor is given by \cite{Free,Baluni,Toimela}

\beq
\Omega^{(0)}= - \frac{N_c}{12 \pi^2}
\left[\mu u(\mu^2-\frac{5}{2}m^2)+
\frac{3}{2}m^4\ln{\left(\frac{\mu+u}{m}\right)}
\right]\;,
\label{Omfree}
\eeq
where $N_c$ is the number of colors and $u\equiv \sqrt{\mu^2-m^2}$.

Using standard quantum field theoretical methods, one obtains 
the complete renormalized exchange energy for a massive quark in the 
$\overline{\hbox{\scriptsize MS}}$ scheme (in the limit $T\rightarrow 0$):
\beq
\Omega^{(1)}=\frac{\alpha_s (N_c^2-1)}{16\pi^3} 
\left[3 \left(m^2 \ln\frac{\mu+u}{m}-\mu u
\right)^2-2 u^4
+m^2\left( 6\ln\frac{\bar{\Lambda}}{m} +4\right) 
\left(\mu u-m^2 \ln\frac{\mu+u}{m}\right)\right]\,.
\label{Om2final}
\eeq

The thermodynamic potential to order $\alpha_s$ for one massive flavor, given 
by the sum of Eqs. (\ref{Om2final}) and (\ref{Omfree}), depends on
the quark chemical potential $\mu$ and on the renormalization subtraction
point $\bar{\Lambda}$ both explicitly and implicitly through the scale
dependence of the strong coupling constant 
$\alpha_s(\bar{\Lambda})$ and the mass
$m(\bar{\Lambda})$.
The scale dependencies of both $\alpha_s$ and $m$, which in the following
we will take to be the mass of the strange quark, are known up to 4-loop
order in the $\overline{\hbox{\scriptsize MS}}$ scheme 
\cite{Vermaseren:1997fq}.
Since we have only determined the free energy to first order in $\alpha_s$, 
we choose
\beq
\alpha_s(\bar{\Lambda})=\frac{4\pi}{\beta_0 L}\left[1-
2 \frac{\beta_1}{\beta_0^2} \frac{\ln{L}}{L}\right] \;\;\;, \;\;\; 
m_s(\bar{\Lambda})=\hat{m}_s\left(\frac{\alpha_s}{\pi}\right)^{4/9}
\left[1+0.895062
\frac{\alpha_s}{\pi}\right] \;,
\eeq
where $L=2 \ln{(\bar{\Lambda}/\Lambda_{\overline{\hbox{\scriptsize MS}}})}$,
$\beta_0=11-2 N_f/3$, and $\beta_1=51-19 N_f/3$ and we take $N_f=3$. The 
scale $\Lambda_{\overline{\hbox{\scriptsize MS}}}$ and the 
invariant mass $\hat{m}_s$ are fixed by requiring \cite{PDB}
$\alpha_s\simeq 0.3$ and $m_s\simeq 100~$MeV at $\bar{\Lambda}=2~$GeV;
one obtains $\Lambda_{\overline{\hbox{\scriptsize MS}}}\simeq 380~$MeV
and $\hat{m}_s\simeq 262~$MeV. With these conventions, the only freedom
left is the choice of $\bar{\Lambda}$. 

To study the effect of the finite strange quark mass on 
the equation of state for 
electrically neutral quark matter with 2 light (massless) 
flavors (up and down quarks) 
and one massive flavor (strange quark), we have to include electrons, with 
chemical potential $\mu_e$ and assume beta equilibrium.
Since neutrinos are lost rather quickly, one may set their chemical
potential to zero, so that chemical equilibrium yields
$\mu_d=\mu_s=\mu$ and $\mu_u+\mu_e=\mu \,$,
with $\mu_u,\mu_d$ and $\mu_s$ the up, down and strange quark chemical
potentials, respectively.
On the other hand, overall charge neutrality requires
$(2/3) n_u-(1/3) n_d-(1/3) n_s-n_e=0 \;,$
where $n_i$ is the number density of the particle species $i$.
Together, the above equations insure that there is only one independent
chemical potential, which we take to be $\mu$.
Number densities are determined from the thermodynamic potential by
$n_i=-(\partial \Omega/\partial \mu)$
and the total energy density is given by
$\epsilon=\Omega+\sum_i \mu_i n_i$,
where $\Omega=\sum_i \left(\Omega_i^{(0)}+\Omega_i^{(1)}\right)$ and 
again $i$ refers to the particle species. The pressure is
$P=n_B \frac{\partial \epsilon}{\partial n_B}-\epsilon \;,$
where $n_B=\frac{1}{3}\left(n_u+n_d+n_s\right)$ is the baryon
number density and the Gibbs potential per particle is given by
$\frac{\partial \epsilon}{\partial n_B}=\left(\mu_u+\mu_d+\mu_s\right)$. 
We restrict the freedom of choice for $\bar{\Lambda}(\mu_u,\mu_d,\mu_s)$ by
requiring that in case of vanishing strange quark mass 
all quark chemical potentials
and number densities become equal so that $P^{(m_s=0)}=-\Omega^{(m_s=0)}$
and, consequently, one has $\mu_e\rightarrow 0$.
Furthermore, in order to compare our findings to existing results in the 
literature \cite{Fraga:2001id,tony}, we require 
that in the massless case $\bar{\Lambda}=2\mu$. Consequently,
we choose $\bar{\Lambda}=\frac{2}{3}\left(\mu_u+\mu_d+\mu_s\right)$,
but have tested that our results are not much affected by other choices
obeying the above conditions.

\begin{figure}[htb]
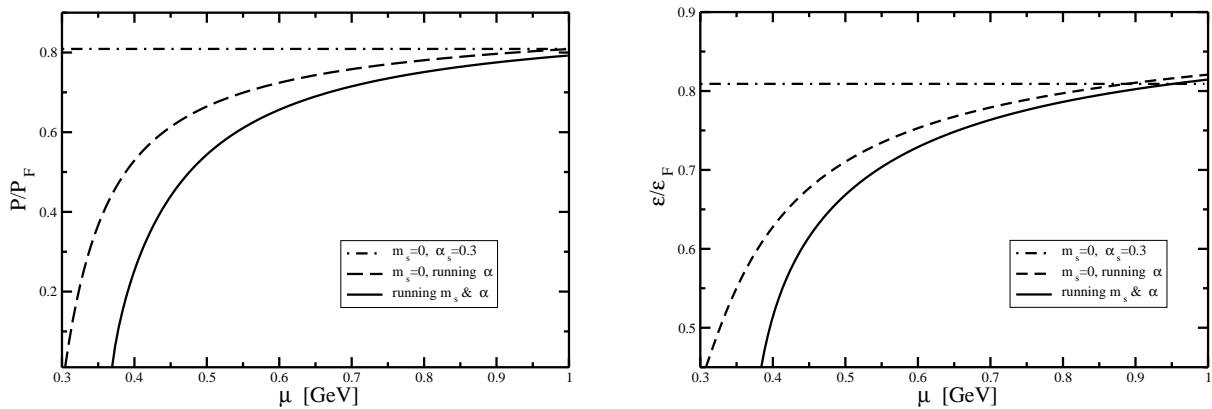

\begin{minipage}[t]{75mm}
\includegraphics[width=\linewidth]{P2comp.eps}
\end{minipage}
\hspace{\fill}
\begin{minipage}[t]{75mm}
\includegraphics[width=\linewidth]{rho2comp.eps}
\end{minipage}
\caption{Pressure and energy density scaled by Stefan-Boltzmann values 
for $\bar{\Lambda}=\frac{2}{3}\left(\mu_u+\mu_d+\mu_s\right)$. 
We show results without renormalization group 
improvements (dash-dotted lines), running coupling (dashed
lines), both for $m_s=0$, and results with running mass and coupling 
(full lines).}
\label{fig2}
\end{figure}

The effects of the finite strange quark mass on
the total pressure and energy density for electrically neutral quark matter
(plus electrons) are given in Fig. \ref{fig2}.
There we show results for 3 light flavors and running coupling, corresponding
to the case considered in \cite{Fraga:2001id}, and for 
2 light flavors and one massive flavor, with both
running coupling and strange quark mass (which reaches $m_s\sim
137$ MeV at $\mu=500$ MeV).

As can be seen from this Figure, there is a
sizable difference between zero and finite strange quark mass pressure
and energy density for the values of the chemical 
potential in the region that 
is relevant for the physics of compact stars. As has been noticed by several 
authors \cite{Fraga:2001xc,andre,alford}, the resulting equation of state, 
$\epsilon=\epsilon(P)$, can be approximated by a non-ideal bag model form 
$\epsilon=4 B_{eff}+ a P$.
Here $a\sim 3$ is a dimensionless coefficient while $B_{eff}$ is 
the effective bag constant of the vacuum. Concentrating on the low-density
part of the equation of state, one finds for massless strange quarks 
the parameters $B_{eff}^{1/4} \simeq 117~$MeV and $a\simeq 2.81$ while the
inclusion of the running mass raises these values to 
$B_{eff}^{1/4} \simeq 137~$MeV and $a\simeq 3.17$ (all values having
been obtained by including a running $\alpha_s$ in the equations of state).
Therefore, we expect 
important consequences in the mass-radius relation of quark stars due to 
the inclusion of a finite mass for the strange quark.

The structure of a quark star is determined by the solution 
of the TOV equations. Corrections to the mass and radius 
of quark stars due to a running strange quark mass can be very 
large, $\sim 25\%$ \cite{Fraga:2004gz}. 

Also, while the most massive star for the
$m_s\!=\!0$ equation of state (with 
$M/M_{\odot}\simeq 3.2$ and radius $\sim 17~$km)
has a central density of $n_B\simeq0.5~{\rm fm}^{-3}$, this number increases
to $n_B\simeq 0.83~{\rm fm}^{-3}$ (at $\mu=470$ MeV) for the heaviest star 
($M/M_{\odot}\simeq 2.16$ at $\sim 12~$km) of the massive equation of state.
The inclusion of $\sim \alpha_s^2$ corrections
to the pressure will increase its non-ideality and produce quark
stars which are smaller, denser and less massive 
\cite{Fraga:2001id,Fraga:2001xc,next}.

My special thanks to my collaborator P. Romatschke for insightful 
comments and for a careful reading of the manuscript. 
I also thank R.D. Pisarski and A. Rebhan for fruitful discussions. 
The work of E.S.F. is partially supported by CAPES, CNPq, FAPERJ 
and FUJB/UFRJ.


\begin{thebibliography}{9}

\bibitem{Fraga:2004gz}
E.~S.~Fraga and P.~Romatschke,
Phys.\ Rev.\ D {\bf 71}, 105014 (2005).

\bibitem{Free}
B.~A.~Freedman and L.~D.~McLerran,
Phys.\ Rev.\  {\bf D16}, 1130 (1977); 
{\it ibid.}, {\bf D16}, 1147 (1977);
{\it ibid.}, {\bf D16}, 1169 (1977);
{\it ibid.},  {\bf D17}, 1109 (1978).

\bibitem{Baluni}
V. Baluni, Phys. Rev. D {\bf 17},  2092  (1978).

\bibitem{Toimela}
T.~Toimela,
Int.\ J.\ Theor.\ Phys.\  {\bf 24}, 901 (1985)
[Erratum-ibid.\  {\bf 26}, 1021 (1987)].

\bibitem{wit}
E. Witten, Phys. Rev. D {\bf 30}, 272 (1984).

\bibitem{bag}
C.~Alcock, E.~Farhi and A.~Olinto, 
Astrophys.\ J.\  {\bf 310}, 261 (1986);
P. Haensel, J.~L. Zdunik, and R. Schaeffer, 
Astron. Astrophys. {\bf 160},  121
(1986).

\bibitem{bod}
A.~R. Bodmer, Phys. Rev. D {\bf 4}, 1601 (1971).

\bibitem{Fraga:2001id}
E.~S.~Fraga, R.~D.~Pisarski and J.~Schaffner-Bielich,
Phys.\ Rev.\ D {\bf 63}, 121702 (2001).

\bibitem{Fraga:2001xc}
E.~S.~Fraga, R.~D.~Pisarski and J.~Schaffner-Bielich,
Nucl.\ Phys.\ A {\bf 702}, 217 (2002).

\bibitem{andre}
A.~Peshier, B.~K\"ampfer and G.~Soff,
Phys.\ Rev.\  {\bf C61}, 045203 (2000); 
Phys.\ Rev.\ D {\bf 66}, 094003 (2002).

\bibitem{tony}
J.~P.~Blaizot, E.~Iancu and A.~Rebhan,
Phys.\ Rev.\ D {\bf 63}, 065003 (2001).

\bibitem{andersen}
J.~O.~Andersen and M.~Strickland,
Phys.\ Rev.\ D {\bf 66}, 105001 (2002).

\bibitem{paul}
A.~Rebhan and P.~Romatschke,
Phys.\ Rev.\ D {\bf 68}, 025022 (2003).

\bibitem{alford}
M.~Alford and S.~Reddy,
Phys.\ Rev.\ D {\bf 67}, 074024 (2003); 
M.~Alford, M.~Braby, M.~Paris and S.~Reddy,
nucl-th/0411016.

\bibitem{next} 
E. S. Fraga and P. Romatschke, in preparation.

\bibitem{Vermaseren:1997fq}
J.~A.~M.~Vermaseren, S.~A.~Larin and T.~van Ritbergen,
Phys.\ Lett.\ B {\bf 405}, 327 (1997).

\bibitem{PDB}
S.~Eidelman {\it et al.}  [Particle Data Group Collaboration],
Phys.\ Lett.\ B {\bf 592}, 1 (2004).
\end{thebibliography}
\end{document}